\documentclass[preprint,5p]{elsarticle}
\usepackage{natbib}
\usepackage{ulem}
\usepackage{graphics}
\usepackage{amssymb}
\usepackage{amsthm}
\usepackage{siunitx}

\biboptions{sort&compress}
\journal{Nuclear Instruments and Methods in Physics Research A }

\begin{document}

\begin{frontmatter}



%\title{Determination of $\beta$ energy-summing in the
%detection of $\beta$-delayed proton emission in double-sided
%silicon-strip detectors}
\title{Exploratory investigation of the HIPPO gas-jet target fluid dynamic properties}


\author[Add1]{Zach Meisel\corref{cor1}}
\ead{zmeisel@nd.edu}
\address[Add1]{Department of Physics,
Joint Institute for Nuclear Astrophysics, University of Notre Dame, Notre Dame, IN 46556, USA} 
\cortext[cor1]{Corresponding author}

\author[Add2]{Ke~Shi}
\address[Add2]{Hessert Laboratory for Aerospace Research, Department of Aerospace and Mechanical Engineering, University of Notre Dame, Notre Dame, IN 46556, USA}

\author[Add2]{Aleksandar Jemcov}

\author[Add1]{Manoel Couder}

\address{}

\begin{abstract}
In order to optimize the performance of gas-jet targets for future
nuclear reaction measurements, a detailed understanding of the
dependence of the gas-jet properties on experiment design parameters is
required. 
Common methods of gas-jet characterization
rely on measuring the effective thickness using nuclear elastic
scattering and energy loss techniques; however, these tests are time
intensive and limit the range of design modifications which can be
explored to improve the properties of the jet as a nuclear reaction target. Thus, a
more rapid jet-characterization method is desired.
We performed the first steps towards
characterizing the gas-jet density distribution of the HIPPO gas-jet
target at the University of Notre Dame's Nuclear Science Laboratory
by reproducing results from $^{20}\rm{Ne}(\alpha,\alpha)^{20}\rm{Ne}$ elastic
scattering measurements with computational fluid dynamics (CFD)
simulations performed with the state-of-the-art 
CFD software {\tt ANSYS Fluent}. 
We find a strong sensitivity to experimental design
parameters of the gas-jet target, such as the jet nozzle geometry
and ambient pressure of the target chamber. 
We argue that improved predictive power will require moving to
three-dimensional simulations and additional benchmarking with
experimental data.
\end{abstract}


\end{frontmatter}

% \linenumbers

\section{Introduction}
\label{Introduction}
Precise $(\alpha,\gamma)$ reaction cross sections are crucial to
predict the nucleosynthesis and nuclear energy generation that
occurs in a variety of stellar and explosive helium-burning
environments~\cite{Magk10,Wies12}. Owing to the relatively low
energies of nuclei in the relevant astrophysical conditions,
traditional nuclear reaction measurements in which an outgoing
$\gamma$-ray is measured often suffer from prohibitively high
$\gamma$-backgrounds~\cite{Caci09}. One approach which has been adopted to overcome this
difficulty employs the so-called recoil separator, in which the
nuclear recoil produced in the $(\alpha,\gamma)$ reaction is
electromagnetically separated from unreacted beam
nuclei and identified with a combination of time-of-flight and
energy-loss measurements~\cite{Ruiz14}. 

The recoil separator technique generally relies on using
inverse kinematics, where a heavy ion beam is impinged on a lighter
nuclear target. As such, $(\alpha,\gamma$) reactions require a
gaseous helium target to be employed. In order to more easily
capture the recoil nuclei, which are emitted from the
target location with some angular
distribution, recoil separators benefit from having a small target
region~\cite{Ruiz14}. To this end, the HIPPO gas-jet
target~\cite{Kont12} was developed to serve as the helium target
for the St.~George recoil separator~\cite{Coud08} at the University
Notre Dame's Nuclear Science Laboratory (NSL).

In order to be suitable as a target for nuclear reaction studies
with St.~George, HIPPO must have a confined (relative to the size of
the ion-beam) region of gaseous helium where the density is as high and
uniform as possible. To achieve these properties, HIPPO produces a
supersonic gas-jet in a windowless, differentially pumped volume.
Though the properties of the gas-jet that
is produced by HIPPO are reported in Reference~\cite{Kont12}, the measurement
techniques used provide indirect information about the gas-jet
volumetric density distribution and are unable to map small-scale properties of the
gas-jet. The ambiguity of the inferred volumetric density
distribution has the potential
to introduce systematic
uncertainties in the absolute cross section determined by nuclear
reaction measurements performed with St.~George.
Additionally, considerable experimental effort
is required to explore even a small number of
design modifications in an effort to improve HIPPO's suitability as a
nuclear reaction target for St.~George. Therefore, a more rapid method of exploring
design modifications is desired.

In order to investigate the properties of the HIPPO gas-jet target,
we have performed computational fluid dynamics (CFD) simulations with the software {\tt
ANSYS Fluent}. We have explored the sensitivity of the gas-jet density
distribution to variations in HIPPO design parameters and
compared our results to experimental data. Our exploratory work
paves the way for future CFD simulations and experimental work which
will eventually
enable the rapid exploration of modifications to the design of HIPPO
and future gas-jets to improve their suitability as nuclear
reaction targets.

We discuss the gas-jet thickness measurements which we compare to
our simulation results in Section~\ref{Measurements}, present our
CFD simulations in Section~\ref{Simulations}, and the method to process
our simulation results to compare to experimental data in
Section~\ref{Comparison}. We discuss our results in
Section~\ref{Discussion} and conclude with closing remarks in
Section~\ref{Conclusions}.

\begin{figure}[ht]
\begin{center}
\includegraphics[width=1.0\columnwidth,angle=0]{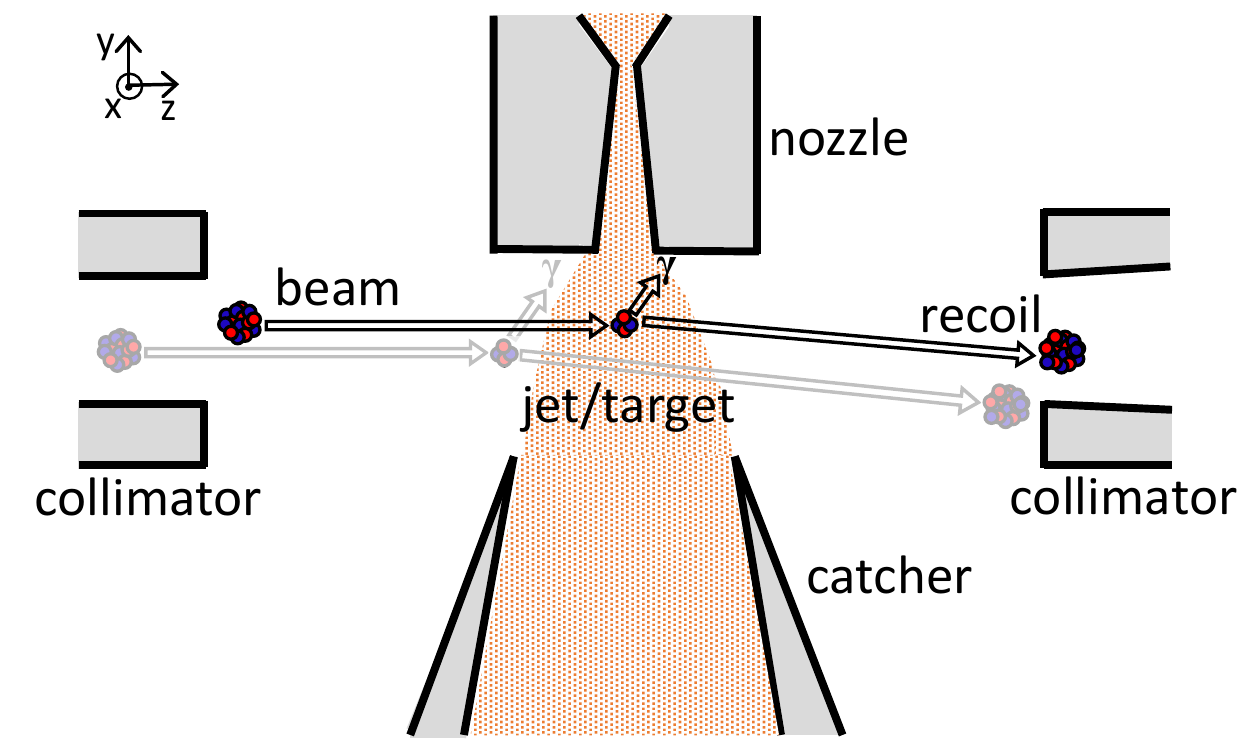}
\caption{
 (color online.) Schematic of the HIPPO gas-jet target, viewed perpendicular to the
 ion beam and gas-jet axes. The opaque (translucent) nuclear
 reaction sequence demonstrates that reactions with the maximum
 expected recoil angle occurring within
 (outside of) the jet-target region are accepted into (rejected
 from) the St.~George recoil separator.
\label{HIPPOcartoon}}
\end{center}
\end{figure}

\section{Gas-jet thickness measurements}
\label{Measurements}

HIPPO is a windowless, supersonic gas-jet contained with
differential pumping that employs recirculating gas
flow~\cite{Kont12}. The
gas-jet, which has an aerial density of $\gtrsim10^{17}$atoms/cm$^{2}$
and a spatial extent of $\sim3\times8$~mm$^{2}$ as seen by the ion
beam, serves as the nuclear
reaction target for the St.~George recoil separator~\cite{Coud08} at
the NSL, as shown schematically in Figure~\ref{HIPPOcartoon}. The jet
is produced by flowing pure gas compressed to a pressure of at least
one bar through an axisymmetric convergent-divergent nozzle into a
chamber with an ambient
pressure of a few millibar, where the underexpanded gas-jet is captured by a
gas-catcher after free-streaming for a distance of $8$~mm. The
central chamber of HIPPO, which contains the gas-jet, is separated
from beam-line vacuum pressures of $\sim10^{-9}$~bar by separately
pumped chambers that are connected along the beam axis by narrow
apertures that accommodate the incoming and outgoing nuclei, but
provide large gas-flow impedances. During reaction measurements, the
central chamber is generally surrounded by a close-packed
$\gamma$-detection
array to detect $\gamma$-rays emitted by the reaction, while a
collimated silicon detector fixed at a forward angle is used to
monitor the incoming ion-beam current.
A full description of the HIPPO gas-jet target and target thickness
measurements is provided in Reference~\cite{Kont12}.

\begin{center}
\begin{table}
\caption{Relative maximum elastic scattering yield (in arbitrary
units) and full-width at half-maximum (FWHM) from a Gaussian fit to yield
profiles measured~\cite{Kont12} for vertical ($y$) deflections toward (+) and
away-from (-) the gas-jet nozzle.  A vertical deflection of $0$~mm
corresponds to the mid-point between the nozzle and catcher, i.e.
4~mm from the nozzle and 4~mm from the catcher. In all cases the
maxima are located along the nozzle-catcher axis, the yield
uncertainty is $\pm0.50$, and the FWHM uncertainty is $\pm$0.20~mm.}
\centering
\begin{tabular}{ c c c }
\hline
$y$-deflection [mm] & Max. Yield [a.u.] & FWHM [mm] \\ \hline
+1.78 & 16.0 & 1.85 \\
+0.89 & 14.3 & 1.90 \\
0 & 12.0 & 2.15 \\
-0.89 & 11.0 & 2.25 \\
+1.78 & 9.8 & 2.30 \\
\end{tabular}
\label{YieldProfileParams}
\end{table}
\end{center}

The HIPPO gas-jet thickness measurement presented in Reference~\cite{Kont12}
employed a 2~MeV $^{20}\rm{Ne}$ beam, produced by the KN Van de
Graaff accelerator at the NSL, impinging on a 1~bar (at the nozzle
inlet) helium gas-jet produced by HIPPO.
A collimated silicon detector, located at the end of an
extension from the central chamber fixed at a
forward angle (See
in Figure~8 of Reference~\cite{Kont12}.), was used to measure elastically-scattered
$\alpha$-particles from
$^{20}\rm{Ne}(\alpha,\alpha)^{20}\rm{Ne}$. 
Normalization of the scattering yield was provided by a Faraday cup
following the gas-jet 
which measured the incoming
$^{20}\rm{Ne}$ beam current. 
A pair of horizontal and vertical electrostatic steerer plates 
located in the beam-line prior to HIPPO 
were used to scan the ion-beam
over different locations on the gas-jet, where collimators
before and after the steerers were used to confine the beam to
a diameter smaller than the gas-jet.
%of $0.8$~mm at the gas-jet location.

The number of elastic scattering events detected
$N_{\rm{det}}$ is the product of the number of
beam particles impinging the jet over some time $N_{\rm{p}}$, the
aerial number density of the jet target atoms in units of
atoms/cm$^{2}$ $N_{\rm{t}}$, the
probability of a beam and jet particle scattering a jet particle at
a given angle $\frac{d\sigma}{d\Omega}(\theta)$, the solid angle
covered by the detection system $d\Omega$, detection efficiency for
the scattered $\alpha$-particle $\epsilon$, and the fraction of the
time in which the data acquisition system was not busy (i.e.
`live') and could acquire new data $f_{\rm{live}}$~\footnote{An additional
correction factor $F$ was included in the form of this equation
presented in Reference~\cite{Beck82} to account for the fact that their
detection set-up would not see all elastically scattered recoils
from their jet.}:
\begin{equation}
N_{\rm{det}}=N_{\rm{p}}N_{\rm{t}}\frac{d\sigma}{d\Omega}(\theta)d\Omega\epsilon
f_{\rm{live}}.
\label{ScatterEqn}
\end{equation}
For the $\alpha$-particles at the measured energies
$\epsilon=1$ for the employed silicon detector, $f_{\rm{live}}$ was
reported by the acquisition system when data was taken, measurements
with radioactive sources and calipers determined
$\theta=64.8^{\circ}\pm0.3^{\circ}$ and
$\Omega=10.0\pm1.4\times10^{-6}$ steradians, and
$\frac{d\sigma}{d\Omega}(\theta)$ was evaluated assuming
Rutherford scattering. The Faraday cup located downstream from the
gas-jet was used to determine the number of beam particles which
impinged upon the gas-jet $N_{\rm{p}}$, where the measured average
charge state ($q=2.3$) of unreacted projectiles exiting the jet was used to
convert between the Faraday cup current and the incoming beam
intensity.

\begin{figure}[h]
\begin{center}
\includegraphics[width=1.0\columnwidth,angle=0]{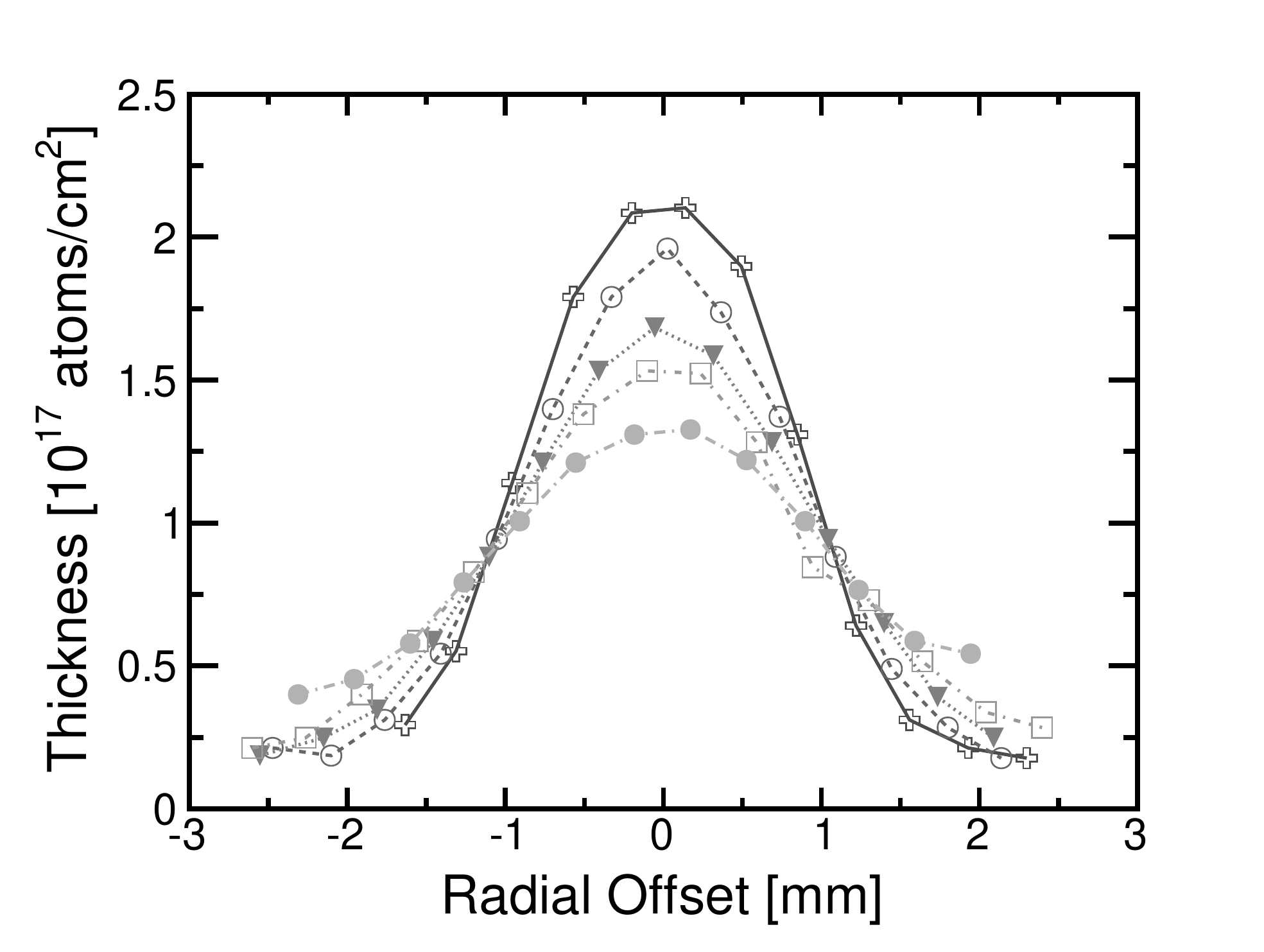}
\caption{Target thickness inferred from the elastic scattering yield from Reference~\cite{Kont12} for a 1~bar (inlet pressure) helium jet from HIPPO
using their optimum nozzle-catcher configuration for a
2~MeV $^{20}\rm{Ne}$ beam impinging on the jet at various
radial and axial positions. The five sets of data, in order of decreasing peak
thickness, correspond to axial distances from the nozzle of 2.22~mm,
3.11~mm, 4.0~mm, 4.89~mm, and 5.89~mm. 
Note that Figure~9 of
Reference~\cite{Kont12}, which presents this data, has a
typographical mistake in the legend, accidentally swapping the
2.22~mm and 3.11~mm cases (``0.89~mm Up" and ``1.78~mm Up" in
Reference~\cite{Kont12}, respectively).
\label{MeasuredYieldProfiles}}
\end{center}
\end{figure}

The ratio of the number of
detected scattering events to the number of incoming projectiles is
the yield $Y$. It is apparent that, since all other quantities are
known and fixed for the duration of the elastic scattering
measurement,
\begin{equation}
Y\propto N_{\rm{t}}.
\label{YieldProportionality}
\end{equation}
Therefore, by measuring the elastic scattering yield $Y$ when
scanning the incoming beam over the spatial extent of the gas-jet,
one measures the spatial distribution of the aerial atomic number density
$N_{\rm{t}}$. For a helium jet with a pressure of 1~bar (for their
final nozzle-catcher geometry), Reference~\cite{Kont12} reports that
their
scattering yield  $Y=12\pm0.5$ corresponds to
$N_{\rm{t}}=(1.68\pm0.14)\times10^{17}$~atoms/cm$^{2}$.
We use
this relationship between the yield and the aerial number density (which is
reported for the `non-deflected' beam, which impinged the jet 4~mm
from the nozzle, centrally along the nozzle-catcher axis) to convert
reported~\cite{Kont12} yields to aerial number densities. 
Yields at various locations along the jet were obtained by adjusting the
voltage of electrostatic steerer plates, where the deflection angle
(and therefore beam-impact position at the jet) was determined by
the steerer plate geometry. The yield was measured for horizontal
(`$x$') and vertical (`$y$') deflections, which correspond to shifts
perpendicular to and along the nozzle-catcher axis, respectively.
The target thicknesses inferred from the measured yields at various 
deflections are shown in
Figure~\ref{MeasuredYieldProfiles}. The properties of Gaussian fits
to the measured yield
profiles are given in Table~\ref{YieldProfileParams}.

\begin{figure}[ht]
\begin{center}
\includegraphics[width=1.0\columnwidth,angle=0]{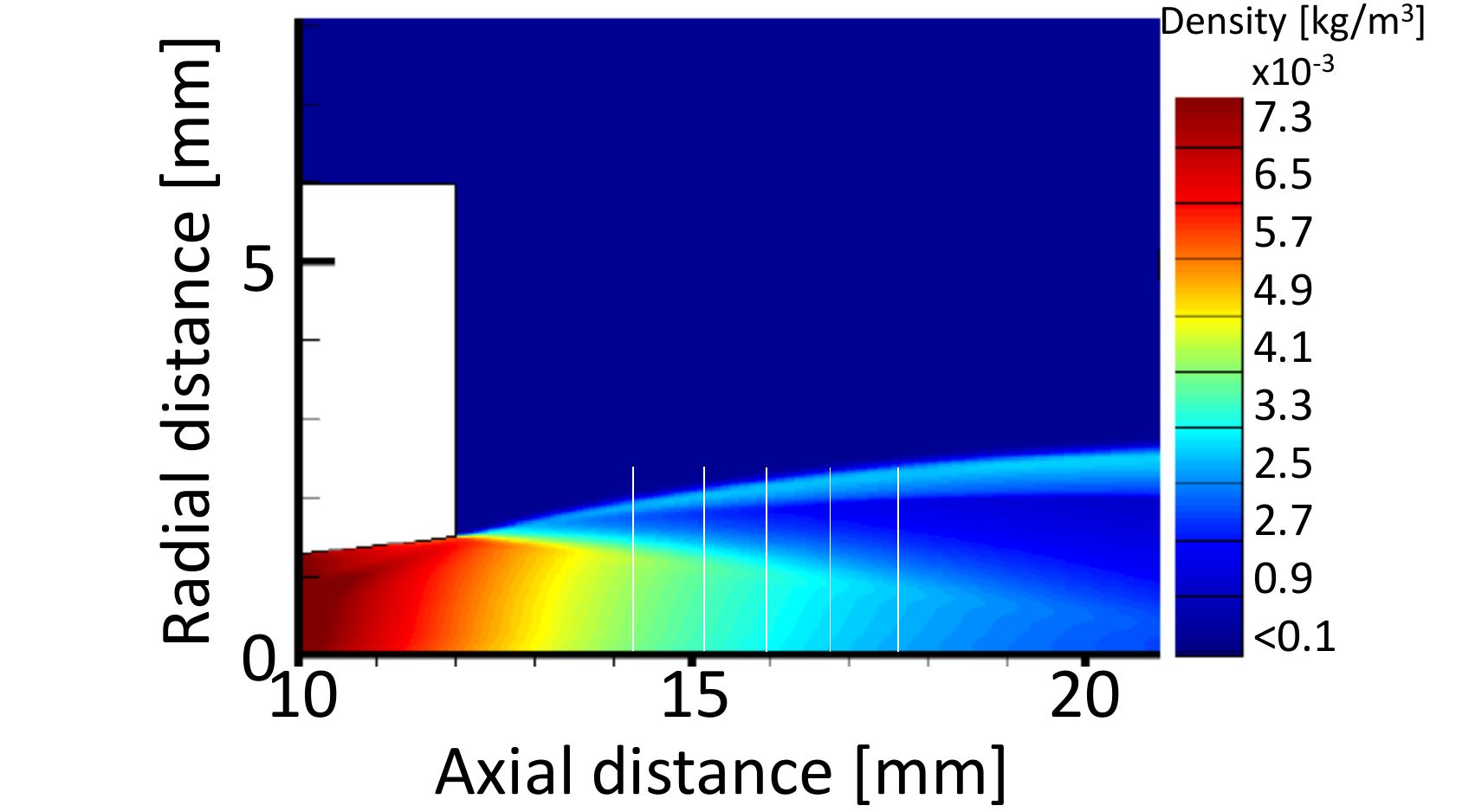}
\caption{
 (color online.) Cross-sectional volumetric mass density profile of the HIPPO gas-jet from an {\tt
 ANSYS Fluent} simulation using a 1~bar inlet pressure and
 $1.5\times10^{-3}$~bar ambient background pressure. The white
 rectangle represents the nozzle and the white vertical lines
 indicate the axial location and radial extent of the yield measurements
 performed in Reference~\cite{Kont12}.
\label{SimulatedJet2D}}
\end{center}
\end{figure}

\section{Gas-jet fluid dynamics simulations}
\label{Simulations}

The state-of-the art computational fluid dynamics (CFD) software
{\tt ANSYS
Fluent}~\footnote{http://www.ansys.com/Products/Fluids/ANSYS-Fluent} (version 14.5.7) was employed to model the
gas-dynamic properties of the HIPPO gas-jet target. 
Rather than simulate the entire `central chamber' region of
Reference~\cite{Kont12}, the simulations modeled the gas-jet within
the nozzle and expanding into an open volume. The omission of the
remaining details of the central chamber, including the upstream and
downstream collimators and gas catcher (each pictured in
Figure~\ref{HIPPOcartoon}), are justified as these
features only serve to modify the central chamber pressure, which is
a pressure boundary condition in the simulations. A fine mesh was used in the
simulations to adequately resolve the spatial properties of the jet.
As an exploratory approach, a two-dimensional axisymmetric
model was adopted in order to reduce computation time. Only one layer of a two-dimensional mesh with
81,300 cells was considered, where the majority of cells were
concentrated near the central jet region. As viscous effects are
expected to be negligible near the gas-jet nozzle
exit~\cite{Pind63,Sala74}, the
inviscid flow assumption was adopted in the simulations for the
purposes of our exploratory study in order to
reduce computation time.

For all simulations a helium jet was modeled using a total pressure
of 1~bar at the nozzle inlet, a 3~mm nozzle-exit diameter, and a
9~mm neck-to-exit length (See Figure~2 of Reference~\cite{Kont12}.).
The nozzle-neck diameter was chosen to be either 1~mm, the nominal
nozzle-neck diameter in Reference~\cite{Kont12}, or 0.72~mm
with an ambient background pressure in the central chamber of either
1.5~mbar or 4.5~mbar to explore the sensitivity of the jet
properties to the experimental design parameters. We were unable to
use the $\sim$0.1~mbar background pressures reported in
Reference~\cite{Kont12} and obtain numerical convergence due to the
rapid dissipation of jet energy at the first downstream (i.e.
increasing in axial distance from the nozzle)
shock~\cite{Garc07}. The impact
of the increased background pressure is discussed in
Section~\ref{Discussion}.
Heat deposition from the
$^{20}\rm{Ne}$ ion beam in the gas-jet was not included, since past
studies have shown it will not affect the jet
properties~\cite{Gorr85}.

The simulated volumetric mass density distribution for the gas-jet in the region
near the nozzle exit to the location of the gas-catcher is shown for
a 1~mm nozzle-neck diameter and 1.5~mbar background pressure in Figure~\ref{SimulatedJet2D}. As
expected for a supersonic expansion of a gas from a high-pressure
nozzle into a low-pressure environment~\cite{Ulbr72}, the density rapidly declines
along the nozzle-catcher
axis with increasing axial separation from the nozzle's narrowest
point. Contrary to suggestions by Reference~\cite{Beck82} that the
volumetric density $\rho$ along the axial direction $y$ should fall as
$\rho\sim1/\sqrt{y}$ following the convergent-divergent nozzle exit,
our simulation results favor the
$\rho\sim1/y^{2}$ behavior they suggest should be more appropriate
inside the nozzle. This is potentially due to the extra dimension in
which gas can expand in our two-dimensional simulations, as opposed
to their~\cite{Beck82} one-dimensional approximation. 

\begin{figure}[ht]
\begin{center}
\includegraphics[width=1.0\columnwidth,angle=0]{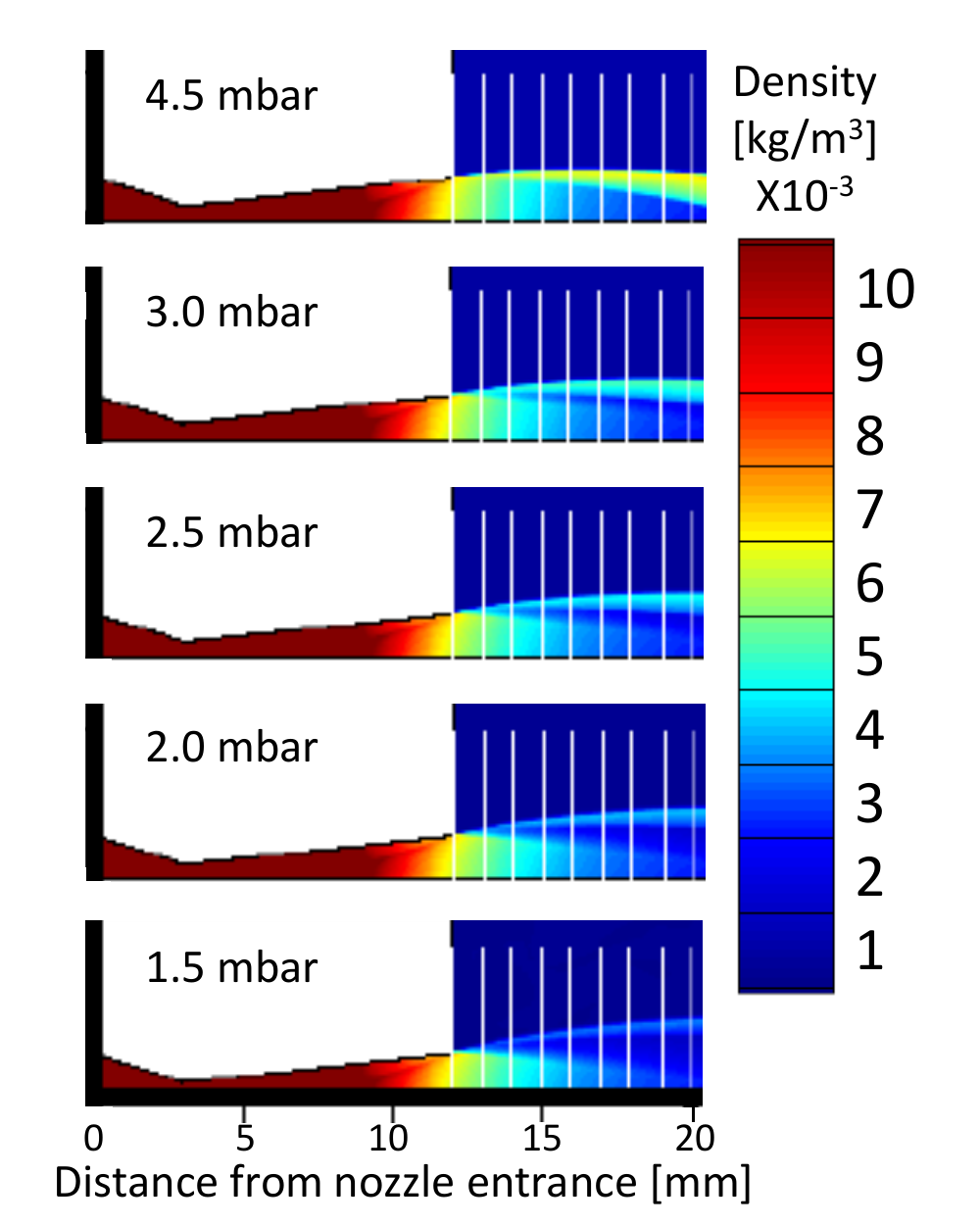}
\caption{
 (color online.) Cross-sectional volumetric mass density profiles of the HIPPO gas-jet from {\tt
 ANSYS Fluent} simulations using a 1~bar nozzle-inlet total pressure for five
 different background pressures. The region shown includes the
 nozzle to the axial distance where the HIPPO gas-catcher would be
 located out to a radial distance of 6~mm. The white vertical lines,
 included for reference, are separated by 1~mm.
\label{Jet2DProfiles5Pressures}}
\end{center}
\end{figure}

\begin{figure}[ht]
\begin{center}
\includegraphics[width=0.8\columnwidth,angle=0]{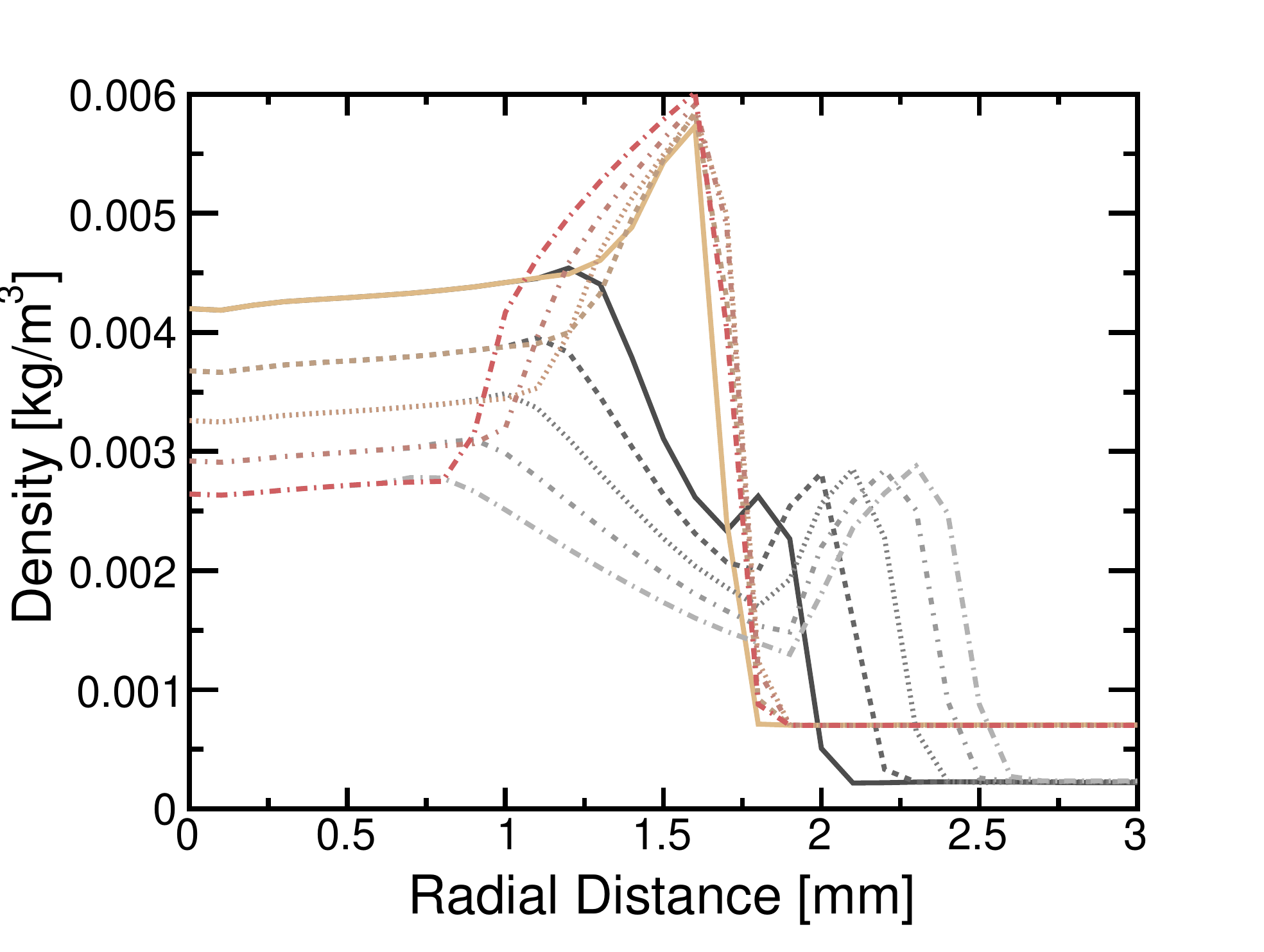}
\caption{
 (color online.) Radial volumetric density profiles of the HIPPO gas-jet from {\tt
 ANSYS Fluent} simulations at axial distances from the nozzle exit of 2, 3, 4, 5, and
 6~mm, in order of decreasing central density for a 1~mm nozzle-neck
 diameter,
 using background pressures of 1.5~mbar
 (gray lines) and 4.5~mbar (colored lines, sharing a maximum density
 at a radial offset of $\sim$1.6~mm). The increased background
 pressure confines the jet to a narrower region, enhancing the
 high-density feature near the jet boundary. Note that the results
 were not normalized for mass-flow.
\label{RadialDensityProfiles}}
\end{center}
\end{figure}

The other striking feature of Figure~\ref{SimulatedJet2D} is
the brief, sharp rise in volumetric density at the outer radial extent of the
gas-jet. This spike in volumetric density is due
to the formation of a shock front by a Prandtl-Meyer expansion fan
emanating from the convex corner at the nozzle neck and reflecting
off of the jet boundary, resulting in compression
waves~\cite{Latv61,Herr68,Chen05,Fran15}. This feature, which is generic to supersonic jets
emerging from a convergent-divergent nozzle due to the strong shear
between the supersonic jet and the rest gas, has been imaged in
similar past studies through electron-beam induced
fluorescence~\cite{Herr68,Bela10} and Schlieren
photography~\cite{Latv61}. We find that increasing the background
pressure within the central chamber enhances the volumetric density increase
near the jet boundary and moves the jet boundary to narrower radii, as shown in
Figures~\ref{Jet2DProfiles5Pressures}
and~\ref{RadialDensityProfiles}. Reference~\cite{Latv61} ascribes the
weakening of the volumetric density enhancement for a decrease in the background
pressure with respect to the nozzle-exit pressure to an increase
in turbulence for these conditions and
therefore a smearing of the boundary where the jet and ambient gas
mix. 

The near-constant volumetric density region of the jet at inner radii is known
as the `potential core'~\cite{Fran15}. As expected, the potential core
becomes narrower with increasing axial distance from the nozzle, as
seen in Figure~\ref{RadialDensityProfiles}.

The radial narrowing of the jet for higher
background pressures (lower nozzle-exit to background pressure
ratios) can be understood by considering the definition of the jet
boundary as the location where the pressure of gas in the jet equals
the ambient gas pressure, as is described in Reference~\cite{Adam59}. As the gas in the jet moves further from
the nozzle it initially expands in area (normal to the
nozzle-catcher axis) and experiences a corresponding decrease in pressure.
Since the jet-boundary is defined as a location of constant
pressure, the decrease in pressure at the boundary of the jet due to
the increase in jet area cannot drive the jet pressure below the
pressure of the ambient background. Therefore, the jet boundary must
turn inward toward the nozzle-catcher axis by some incremental angle while progressing from the nozzle
to the first downstream (with respect to the nozzle exit) shock
(`Mach disk'), which is located at axial distances larger than those shown
in Figures~\ref{SimulatedJet2D} and~\ref{Jet2DProfiles5Pressures}. For higher background
pressures, the pressure of the jet is driven down to the ambient pressure
by the increase in jet area at closer axial distances to the nozzle.
Thus, for relatively higher background pressures, the
supersonic gas-jet is narrower, as seen in
Figures~\ref{Jet2DProfiles5Pressures}
and~\ref{RadialDensityProfiles}. For the lower ambient pressure
case, the jet-boundary is still expanding to larger radii at the
axial distances from the nozzle which are shown in
Figure~\ref{RadialDensityProfiles}, whereas the higher background
pressure case reaches its apex in jet-boundary radius near this
axial distance. Thus, the location of the jet boundary is roughly
constant for the 4.5~mbar ambient pressure case.

For a fixed mass-flow, reducing the nozzle-neck diameter has the somewhat trivial impact of 
reducing the overall jet density. However, this would be recovered in
the experimental set-up by increasing the gas charge in the system.
The reduced jet-density, and therefore reduced ratio between the jet
and ambient pressures, causes a narrowing of the jet, which is
counteracted by the increased opening angle of the
nozzle~\cite{Latv61}. The combined impact on the jet aerial density profile is shown in
Figure~\ref{BeamShapeEffect}.

\section{Comparison between data and simulations}
\label{Comparison}

In order to obtain a detailed comparison between the {\tt ANSYS
Fluent} simulation results and experimental data, we convert the
simulated volumetric mass density distributions into distributions
of aerial number density $N_{\rm{t}}$.

This is accomplished by evaluating
the integral
\begin{equation}
N_{{\rm{t}}{,y}}=\int n_{{\rm{t}}{,y}}(\ell)d\ell
\label{NumberIntegralBasic}
\end{equation}
along a path through the jet $\ell$, where the volumetric number density
distribution $n_{{\rm{t}},y}(\ell)$ (atoms/cm$^{3}$) from the simulation results for a given
axial offset $y$ is used. Since the nozzle is axisymmetric, so too
is the jet structure and therefore our jet density distribution is
described by a radial function of the volumetric mass density
$\rho_{y}(r)$. When
integrating along the longitudinal direction $x$, in which the ion beam
traverses the gas-jet, a polar coordinate transformation
is used to obtain a radial distance $r$ from
the nozzle-catcher axis and therefore a volumetric density at that location. When
integrating the jet thickness, locations with pressure below the
ambient background pressure are ignored. This is because the ambient
pressure in the experimental conditions is at least ten times lower
the possible background pressures employed in the simulations, due to numerical convergence
issues for too large a pressure range within a given simulation.
Therefore, scattering of the ion beam off of ambient gas in the
thickness measurement should have been negligible.
The volumetric number density distribution $n_{\rm{t}}(x,y)$ (where
$x$ and $y$ are used due to the Cartesian nature of the simulation
grid) is trivially obtained
from the simulation volumetric mass density distribution via the relationship
$\rho=\frac{n_{t}m_{\rm{mol}}}{N_{\rm{A}}}$, where $m_{\rm{mol}}$ is
the molar mass of helium and $N_{\rm{A}}$ is Avogadro's number.

An additional correction must be made to account for the finite
size of the ion beam as it impinges on the gas-jet.  To account for
this, we calculate an effective target thickness at each point
$N_{\rm{t}}^{\rm{eff}}(x',y')$ by
assuming a beam distribution of width $\Delta_{x}$ and weighting the local target
thickness by the neighboring target thicknesses $N_{\rm{t}}(x+\delta
x,y)$, where the weight $w(x+\delta x)$ is
determined by the beam distribution shape.
The resultant effective target thickness is
\begin{equation}
N_{\rm{t}}^{\rm{eff}}(x',y')=\int_{x'-\Delta_{x}/2}^{x'+\Delta_{x}/2}w(x)N_{\rm{t}}(x,y')dx.
\label{EffectiveThicknessAtXY}
\end{equation}

\begin{figure*}[]
\begin{center}
\includegraphics[width=1.5\columnwidth,angle=0]{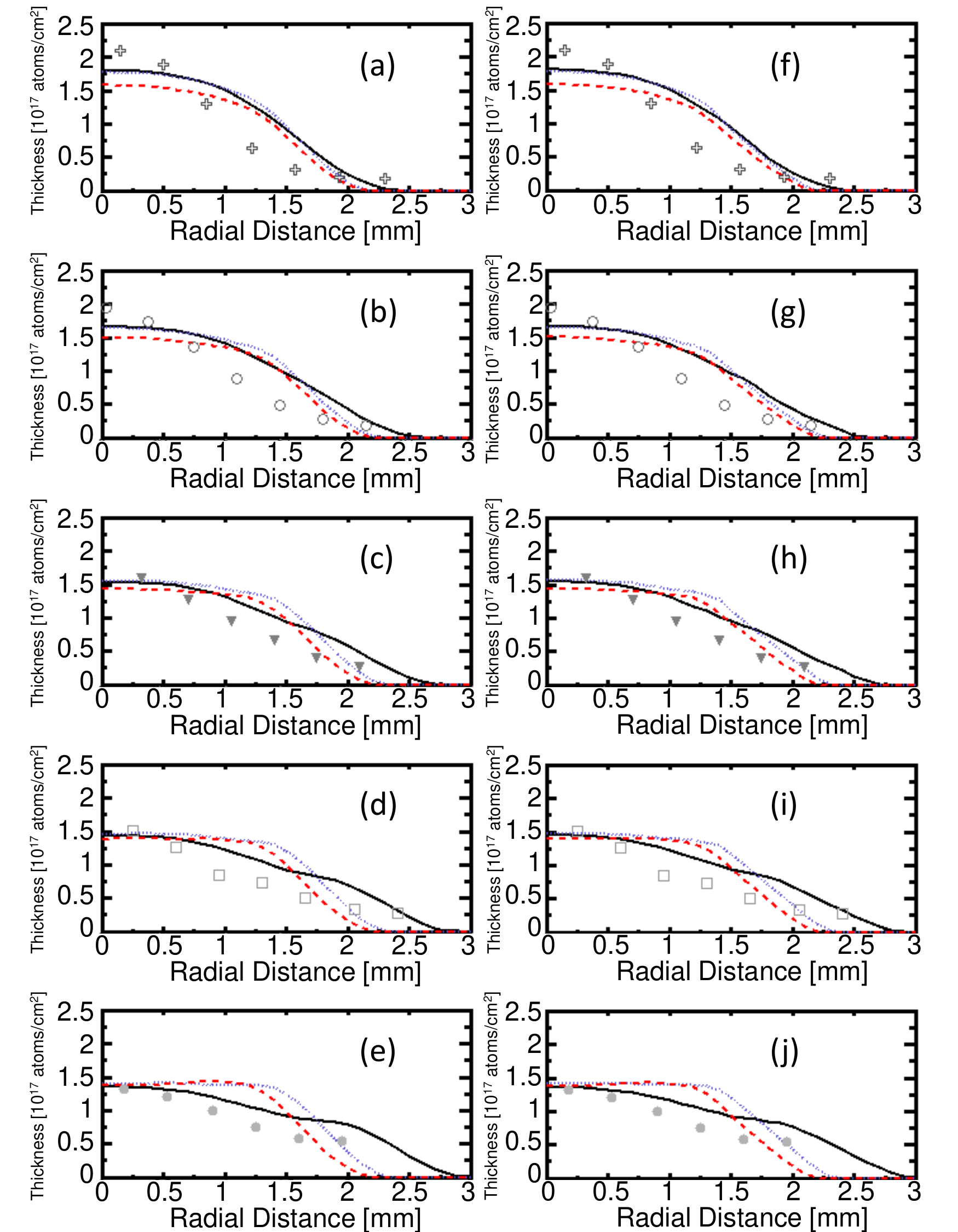}
\caption{
 (color online.) Comparison between the measured (points) and simulated (lines)
 aerial density (`thickness') for a scan over the radial profile of the
 gas-jet at axial distances of $\sim2-6$~mm (top to bottom in
 $\sim1$~mm
 increments) from the nozzle exit. The
 simulation results in the left column assume a Gaussian
 distribution of the ion beam used for the elastic scattering
 measurement with a full-width at half-maximum of 0.8~mm, 
 while the simulation
 results in the right column assume a uniform ion beam distribution with a
 width of 0.8~mm. The simulation conditions for the solid black,
 dashed red, and
 dotted blue curves are 
 $d_{\rm{nozzle-neck}}=1$~mm $p_{\rm{background}}=1.5$~mbar,
 $d_{\rm{nozzle-neck}}=1$~mm $p_{\rm{background}}=4.5$~mbar, and
 $d_{\rm{nozzle-neck}}=0.72$~mm $p_{\rm{background}}=1.5$~mbar,
 respectively. The magnitude of the three simulation results have
 been separately scaled ($\times$ 0.77, 0.66, and 1.5 for the black, red, and
 blue curves, respectively) to roughly match the measured thickness at the
 nozzle-catcher axis. Uncertainties for experimental data are
 obscured by the size of the data points.
\label{BeamShapeEffect}}
\end{center}
\end{figure*}

Measurements reported in
Reference~\cite{Kont12} determined the beam width was 0.8~mm,
however, they neglected to note the beam distribution shape.
Therefore, we have separately employed a uniform weighting
distribution 0.8~mm wide and a Gaussian weighting distribution with
a full-width at half-maximum of 0.8~mm. As shown in
Figure~\ref{BeamShapeEffect}, we found little sensitivity to our
choice of weight for the beam distribution. We do not perform a
weighting over the axial dimension $y$, since, as seen in
Figure~\ref{RadialDensityProfiles}, the increase and decrease in jet
volumetric density for distances closer to and further from the nozzle exit are
nearly equivalent and therefore essentially counteract each other.
This was confirmed with a simple step-function weighting for a low
resolution of axial distances.

The effective target thickness distributions derived from our {\tt ANSYS
Fluent} simulation results are compared to experimental data (shown
separately in Figure~\ref{MeasuredYieldProfiles}) in Figure~\ref{BeamShapeEffect}.
The simulation results shown are for the axial distance at the nearest whole millimeter to
the experimental results and assume two different nozzle-neck diameters
and two different ambient background pressures. The simulation results have been
scaled for three representative simulations which we have chosen to highlight, with three different scaling
factors to roughly reproduce the measured central target thickness.
The unscaled simulation results reproduced the measured central thickness
within 50\%.

\section{Discussion}
\label{Discussion}

The radial target thickness profiles for various axial distances
from the nozzle show qualitative agreement with the data (See
Figure~\ref{BeamShapeEffect}.) for each of the assumed ion beam
distribution shapes and experiment design parameters. While the
overall width of the gas-jet is reproduced, we find a more
plateau-like radial behavior, resulting in a systematic
overestimation of jet thickness near the half-width of the jet. This
effect is less pronounced for simulations employing the nominal
experimental nozzle-neck diameter and lowest ambient background
pressure. Therefore, we expect the discrepancy at the half-width of
the jet can partially be explained by our use of ambient background
pressures that exceeded the experimentally measured value, which was
required to achieve numerical convergence (See
Section~\ref{Simulations}.). We surmise that increased mesh
resolution around the regions containing large pressure gradients
may remedy our numerical convergence issue, at the cost of
computation time.

The remaining discrepancy between the data and simulation results
is likely due to the two-dimensional treatment of a
three-dimensional phenomenon. Employing two-dimensions enforces an
artificial symmetry that results in a systematic deviation of the
simulated jet behavior from reality, for instance, shifting the
location of the Mach disk (beyond the axial distances pictured in
Figures~\ref{SimulatedJet2D} and~\ref{Jet2DProfiles5Pressures}) for all nozzle-exit/ambient pressure
ratios~\cite{Wilk06,Cina06}. The deviation between simulated and measured
jet densities is particularly severe for increased
nozzle-exit/ambient pressure ratios~\cite{Garc07}. It is interesting
to note that hydrodynamic simulations of core collapse supernovae
also have markedly different results for 2D and 3D simulations, in
that case due to the inverse behavior of energy cascading between
different length scales via turbulence~\cite{Jank16}. In the present
case, turbulent behavior within the nozzle may not be captured
properly since our assumption of inviscid flow begins to break down for the
relatively small Reynolds number corresponding to a millimeter-size
nozzle with supersonic flow~\cite{Loui14}.

As seen in Figure~\ref{RadialDensityProfiles}, 
the qualitative impact of reducing the ambient background pressure is a
broadening of the gas-jet and a lessening of the sharp volumetric density
increase present near the jet boundary. These effects conspire to
erase the plateau-like nature of the jet-thickness profile,
resulting in a smoother transition from the central thickness to the
ambient background conditions. Interestingly, modifying the
neck-nozzle to a more narrow diameter while keeping the low ambient
background pressure reinstates the plateau-like feature.

A comparison between the left and right columns of
Figure~\ref{BeamShapeEffect} demonstrates an encouraging lack of
sensitivity to our assumptions about the spatial distribution of the ion
beam used to perform the thickness measurement of the jet via
elastic scattering. We therefore conclude that our gas-jet behavior
is primarily sensitive to design parameters of the gas-jet target
system, in particular, the features impacting the ambient pressure of
the target chamber. Nonetheless, future studies of gas-jet properties via
elastic scattering would benefit from a measurement of the ion beam
profile properties, beyond only the `width'.

Contrary to prior expectations~\cite{Kont12}, we find the expected
radial volumetric density distribution for the HIPPO gas-jet differs markedly
from the profile na\"{i}vely deduced from the elastic scattering
yield data.
We find the jet-density distribution
does not resemble a Gaussian distribution. To the contrary, the
volumetric density is rather constant near the nozzle-catcher axis and,
following a period of decreasing density, increases again near the
jet boundary. For relatively higher ambient pressures, the initial
decrease in volumetric density in the radial direction is absent altogether. The substantial
change in the jet-density distribution for increased background
pressures is an important finding, as the introduction of a high
atomic number gas into the central chamber is being considered to
ensure charge-state equilibration of recoil ions exiting the
gas-target, as has been done previously~\cite{Schu04}.

The enhancement in volumetric density near the jet boundary for 
increased background pressures may explain the reduced gas-catching
efficiency observed by Reference~\cite{Kont12} when argon is
introduced into the HIPPO central chamber, since a larger fraction
of the jet gas would flow towards the outer edge of the catcher (See
Figure~\ref{Jet2DProfiles5Pressures}.). Counterarguments to this
observation are that the catcher radius used in
Reference~\cite{Kont12} is larger than the radial extent of all jets
simulated in this work and that, rather than narrow as expected, their measured
jet distribution slightly broadened. We note that their result is
somewhat counterintuitive since, as discussed in
Section~\ref{Simulations}, arguments from first principles suggest
that increased ambient pressures should result in a narrower jet
boundary~\cite{Adam59}. It is interesting to note that few numerical
or experimental studies have been performed to date in order to determine the
diameter of an underexpanded gas jet~\cite{Fran15}.

The case could be made that the details of the volumetric density
distribution $n_{\rm{t}}(\ell)$ of the gas-jet are inconsequential, as it is the aerial number
density (`target thickness') $N_{\rm{t}}$
that matters for a nuclear reaction cross section
measurement~\cite{Rolf88}. 
However, the radial (with respect to the nozzle-catcher axis) density distribution of
the gas-jet will be important to consider when optimizing HIPPO as a
target for the St.~George recoil separator. Due to the ion-optical
design of St.~George, the ability of the separator
to accept recoil nuclei emitted from nuclear reactions is
compromised by shifts of the target location from the assumed
target region~\cite{Coud08}, as illustrated in
Figure~\ref{HIPPOcartoon}. If a substantial fraction of recoil nuclei were emitted
from non-optimal positions, as would be the case if the target gas
was primarily located at large radii from the nozzle-catcher axis,
then a drastic underestimation of the deduced nuclear reaction cross
section could occur. Furthermore, an extended target depth increases
the effective recoil spot size at the target location, since the
trajectories of ion optical rays are determined by the planar position and
angle at the target center. This effect increases the width of
recoils at the St.~George mass-rejection slits, adversely affecting the
beam-rejection~\cite{Coud08}.

Future studies will be critical to optimizing the
performance of HIPPO as a nuclear reaction target for the St.~George recoil
separator. The experimental design parameters, e.g. nozzle
properties, should be explored
in a systematic study with {\tt ANSYS Fluent}, including exploration of
nozzles other than the convergent-divergent type and 
non-axisymmetric designs. This rapid prototyping can then be
accompanied by the more time-consuming experimental verification
studies of the gas-jet density distributions resulting from
the most promising simulations. 
Given the ambiguity demonstrated in this work of elastic scattering
yield data with respect to determining the gas-jet volumetric density profile,
future verification studies should include alternative methods such
as electron-beam induced fluorescence and Schlieren
photography, if possible~\cite{Beck82,Fave15}. The findings of this and
future studies of the fluid dynamic properties of the HIPPO gas-jet
target will provide important input to help improve the performance
of future gas-jet nuclear reaction targets required for recoil
separators~\cite{Chip14}, reaction studies in ion storage
rings~\cite{Grie12,Mei15}, and reaction studies requiring high
beam-intensities~\cite{Fave15}, which will play a pivotal role in
advancing experimental nuclear astrophysics.

\section{Conclusions}
\label{Conclusions}

In summary, we have performed state-of-the-art CFD simulations with
the software {\tt ANSYS Fluent} of the HIPPO gas-jet target and
compared our results to $^{20}\rm{Ne}(\alpha,\alpha)^{20}\rm{Ne}$
elastic scattering measurements of the target thickness. We find
qualitative agreement between our simulation results and
experimental data. Our results demonstrate a strong sensitivity to
the design conditions of the gas-jet target, highlighting the need
for a systematic exploration of modifications of the nominal design
in order to improve the performance of HIPPO as a nuclear
reaction target. However, we note that the strength of our
conclusions is limited by approximations presently employed in our
simulations. Specifically, future CFD studies should explore the
impact of moving from 2D to 3D and including viscous flow. More
robust comparisons between simulations and data may be obtained by
using redundant, alternative mechanisms for probing the jet
volumetric density
distribution, such as
electron-beam induced fluorescence and Schlieren photography.
The results obtained in this exploratory investigation are a
crucial first step towards optimizing HIPPO as a nuclear
reaction target for the St.~George recoil separator. Future studies
of the HIPPO jet properties
will not only benefit the performance of HIPPO, but also gas-jet
targets employed around the world for future nuclear reaction
studies.

\section{Acknowledgements}
\label{Acknowledgements}
This material is based upon work supported by the National Science
Foundation under Grants No. 1419765 and 1430152. We thank
the University of Notre Dame Turbomachinery Laboratory for their
support.
%NSF grants are NSL (starting mid-2014) and JINA-CEE (starting fall 2014)


%% References


\bibliographystyle{model1a-num-names}
\bibliography{HIPPOreferences}





\end{document}